\documentclass{article}
\usepackage{spconf,amsmath, amssymb, amsfonts, hyperref}
\usepackage{textcomp}
\usepackage{xcolor}
\usepackage{array}
\usepackage{booktabs}
\usepackage{multirow}
\usepackage{tabularx}
\usepackage{makecell}
\usepackage[table]{xcolor}
\usepackage[dvipsnames]{xcolor}
\usepackage{graphicx}
\usepackage{bm}

\title{PENGUIN: General Vital Sign Reconstruction from PPG \\with Flow Matching State Space Model}
\name{Shuntaro Suzuki\thanks{Our code is availalbe at \href{https://github.com/Neurogica/PENGUIN}{https://github.com/Neurogica/PENGUIN}},\quad Shuitsu Koyama,\quad Shinnosuke Hirano,\quad Shunya Nagashima}
\address{Neurogica Inc.}

\begin{document}
%
\maketitle
\begin{abstract}
Photoplethysmography (PPG) plays a crucial role in continuous cardiovascular health monitoring as a non-invasive and cost-effective modality.
However, PPG signals are susceptible to motion artifacts and noise, making accurate estimation of vital signs such as arterial blood pressure (ABP) challenging.
Existing estimation methods are often restricted to a single-task or environment, limiting their generalizability across diverse PPG decoding scenarios. 
Moreover, recent general-purpose approaches typically rely on predictions over multi-second intervals, discarding the morphological characteristics of vital signs. 
To address these challenges, we propose PENGUIN, a generative flow-matching framework that extends deep state space models, enabling fine-grained conditioning on PPG for reconstructing multiple vital signs as continuous waveforms.
We evaluate PENGUIN using six real-world PPG datasets across three distinct vital sign reconstruction tasks (electrocardiogram reconstruction, respiratory monitoring, and ABP monitoring). 
Our method consistently outperformed both task-specific and general-purpose baselines, demonstrating PENGUIN as a general framework for robust vital sign reconstruction from PPG.

\end{abstract}

\section{Introduction}
\label{sec:intro}
Cardiovascular disease (CVD) is one of the leading causes of mortality and morbidity worldwide, with an estimated 1.28 billion people affected by hypertension, a major risk factor for CVD~\cite{Fuchs, WHO}.  
Despite its prevalence, many individuals remain undiagnosed or unaware of their condition, highlighting the critical need for continuous monitoring of relevant vital signs (e.g., heart rate and arterial blood pressure; ABP) to enable early detection of CVD.
To this end, a variety of measurement techniques targeting cardiovascular metrics have been explored, including electrocardiography (ECG), impedance pneumography, and cuff-based sphygmomanometry.

Among various modalities, photoplethysmography (PPG), an optical technique measuring blood flow in the microvascular bed of tissue, has attracted considerable attention as a non-invasive, cost-effective modality particularly well-suited for continuous, long-term monitoring of cardiovascular health~\cite{charlton, PPGreview}.
Its applications span a wide range, from clinical pulse oximeters to consumer health devices such as smartwatches and smart rings.
However, compared to other monitoring techniques, PPG signals are highly susceptible to noise and motion artifacts~\cite{Alfonso, Bent}, and their quality is markedly influenced by individual factors such as skin tone and body composition~\cite{Puranen,Ajmal}.
These limitations make the accurate estimation of vital signs particularly challenging.

Therefore, various PPG decoding methods have been explored~\cite{CycleGAN,CardioFlow,RespDiff}. 
However, most of these models are tailored to a single task or environment, limiting their generalizability across diverse PPG decoding scenarios.
Although a few recent models, such as PaPaGei~\cite{PaPaGei}, have been proposed to predict multiple vital signs from PPG, they rely on discrete predictions at multi-second intervals, limiting their ability to capture the morphological characteristics of vital signs (e.g., LF/HF ratio from ECG and vascular stiffness from ABP).

In this study, we propose PENGUIN, a model capable of reconstructing multiple vital signs from PPG as continuous waveforms. 
This enables fine-grained monitoring of vital signs, including their morphological characteristics.
PENGUIN leverages the Optimal Transport Conditional Flow Matching (OT-CFM) framework~\cite{OT-CFM}, enabling high-quality generation of vital signs directly from their corresponding PPG signals with only a few sampling steps. 
Furthermore, by extending deep state space models (SSMs)~\cite{S5, Mamba}, we introduce a dual-stream Flow-SSM block. 
This architecture explicitly models PPG and vital signs in parallel, thereby enabling fine-grained PPG conditioning.
Experimental results demonstrate that the proposed model outperforms previous works, including task-specific models, across six PPG datasets covering the reconstruction of three distinct vital signs (ECG, respiratory waveform, and ABP). 
To the best of our knowledge, this is the first study to address the continuous reconstruction of multiple vital signs from PPG.
\section{Related Works}
\begin{figure*}
    \centering
    \includegraphics[width=0.73\linewidth]{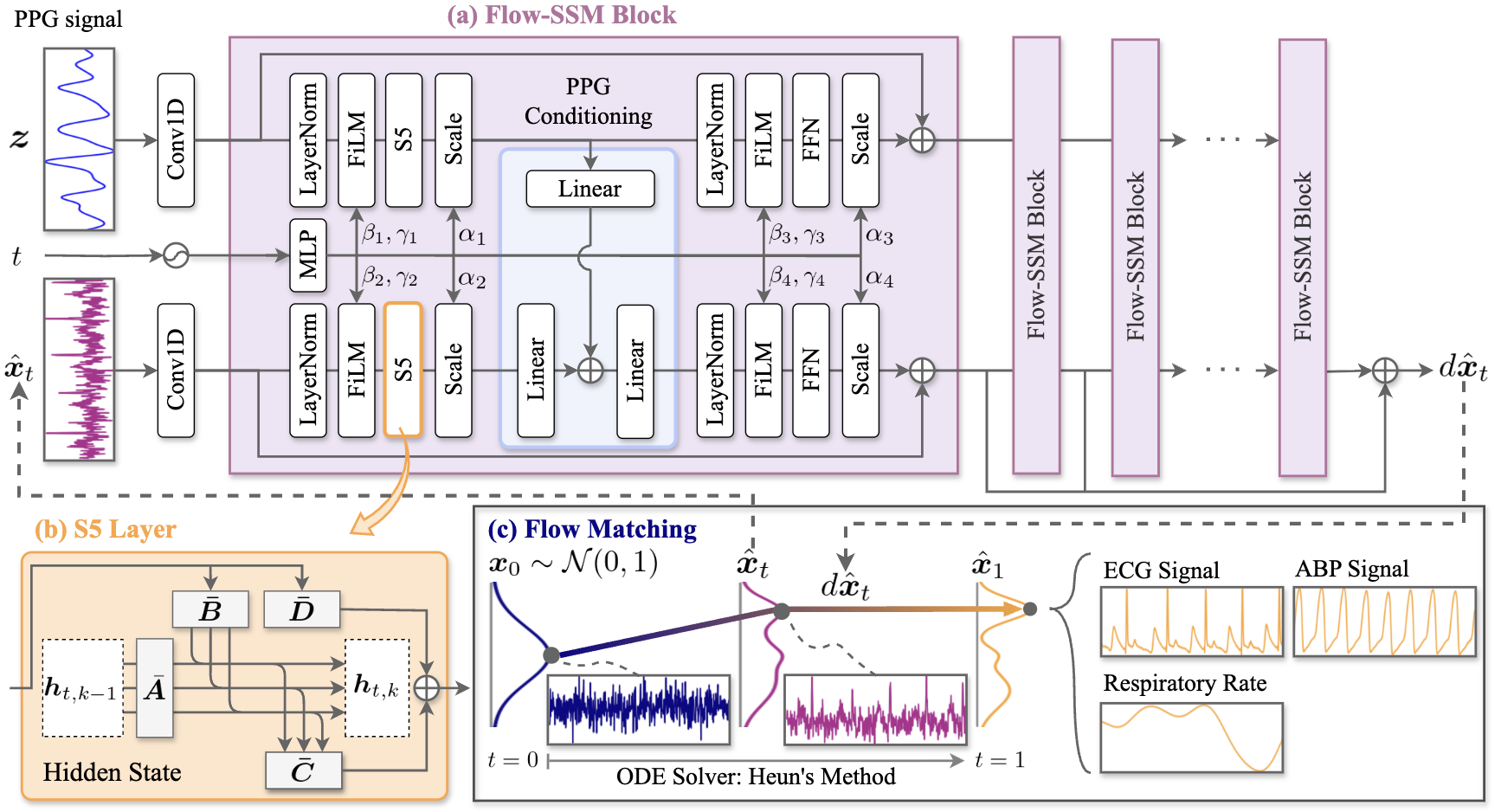}
    \caption{
    Framework of \textbf{PENGUIN}. (a) A stack of Flow-SSM blocks outputs the derivative of the vital sign flow, conditioned on PPG. (b) S5 layer~\cite{S5}, a variant of SSMs, is extended for sequence modeling. (c) PENGUIN is built upon the flow matching framework~\cite{OT-CFM}.
    }
    \label{fig:model_arch}
\end{figure*}

PPG signals are modulated by various vital signs, and deep learning-based approaches for vital sign estimation have been extensively investigated, as summarized in \cite{Naeini, PPGreview}. 
A substantial body of work relies on CNNs, demonstrating their effectiveness across diverse tasks, including heart rate estimation~\cite{Kechris, Song} and ABP monitoring~\cite{Zhang, Chen}.

Despite these successes, prior studies face two key limitations.
First, many studies are restricted to a single task or environment, limiting their generalizability to broader PPG decoding scenarios. 
To address this, a few recent works~\cite{Abbaspourazad, PaPaGei} have designed a model capable of predicting multiple vital signs from PPG. 
Second, previous approaches predict vital signs in a discrete manner at fixed intervals of several seconds, thereby discarding the morphological characteristics inherent in waveforms. 
To overcome this limitation, recent studies such as RDDM~\cite{RDDM} and RespDiff~\cite{RespDiff} reconstruct vital sign waveforms directly by leveraging generative frameworks~\cite{GAN, DDPM}.
Nevertheless, existing approaches still fall short of addressing both challenges simultaneously (i.e., developing a model capable of reconstructing multiple vital sign waveforms from PPG), leaving a significant gap in the field.

\section{Proposed Method}
We propose PENGUIN, a generative flow matching framework capable of reconstructing multiple vital sign waveforms (e.g., ECG) from noise, conditioned on PPG.
Fig.~\ref{fig:model_arch} shows the overall framework of our proposed method. 

\subsection{Flow Matching Formulation}
We begin by deriving the basic principles of OT-CFM~\cite{OT-CFM}, upon which our model is built.
Continuous Normalizing Flows (CNFs)~\cite{NeuralODE} such as OT-CFM constitute a family of generative models that learn conditional mappings from a prior distribution $\bm{x}_0 \sim p_0$ to a data distribution $\bm{x}_1 \sim p_1$.
In our method, we adopt $p_0 = \mathcal{N}(0,1)$ and condition on PPG signals $\bm{z}\in\mathbb{R}^K$, with the goal of mapping to the distribution $p_1$ of vital signs over $\mathbb{R}^K$.
Here, $K$ denotes the sequence length of the PPG signal and the corresponding vital signs.

In CNFs, we consider a probability path $(p_t)_{0 \leq t \leq 1}$ that pushes $p_0$ towards $p_1$ (a.k.a. flow).
The gradient of $\bm{x}_t\sim p_t$ is modeled by a time-dependent velocity field $\bm{u}:[0,1]\times\mathbb{R}^K\times\mathbb{R}^{K} \rightarrow \mathbb{R}^K$, defined through the following ODE:
\begin{align}
d\bm{x}_t = \bm{u}_t(\bm{x}_t | \bm{z}).
\end{align}
By parameterizing the velocity field with a neural network $\bm{u}_t^\theta$, the CNF objective, known as the flow matching loss, can be expressed as the following regression:
\begin{align}
  \label{eq:FM_loss}
  \mathcal{L}_{\textrm{FM}}(\theta) = \mathbb{E}_{t,p_t(\bm{x}_t)}||\bm{u}_t^\theta(\bm{x}_t|\bm{z}) - \bm{u}_t(\bm{x}_t|\bm{z})||^2.
\end{align}
However, in Eq.~\ref{eq:FM_loss}, the distribution $p_t$ is unknown, making it intractable to directly derive $\bm{u}_t$.
As an alternative, CNFs employ the conditional flow matching loss, defined as
\begin{align}
    \label{eq:CFM_loss}
  \mathcal{L}_{\textrm{CFM}}(\theta) = \mathbb{E}_{t,p_t(\bm{x}_t|\bm{x}_1),p_1(\bm{x}_1)}||\bm{u}_t^\theta(\bm{x}_t|\bm{z}) - \bm{u}_t(\bm{x}_t|\bm{x}_1, \bm{z})||^2.
\end{align}
Eq.~\ref{eq:CFM_loss} has been proven to have an identical gradient as Eq.~\ref{eq:FM_loss} w.r.t. $\theta$~\cite{OT-CFM}.
In OT-CFM, the probability path from $p_0$ to $p_1$ is learned as the shortest transport path, enabling high-quality sampling even with a small number of sampling steps. In this formulation, the flow is given by $\bm{x}_t = (1-t)\bm{x}_0 + t\bm{x}_1$, and the corresponding velocity field is $\bm{u}_t(\bm{x}_t \mid \bm{x}_1,\bm{z}) = \bm{x}_1 - \bm{x}_0$.

After training $\bm{u}_t^\theta$, an arbitrary ODE solver can be used to generate the synthesized vital sign $\hat{\bm{x}}_1\in\mathbb{R}^K$ from $\bm{x}_0$ (Fig.~\ref{fig:model_arch}(c)). In this work, we adopt Heun’s method~\cite{ODE}, which has shown promising results in generative sampling~\cite{ConsistencyModels}.

\subsection{Deep State Space Model Formulation}
To model PPG and vital sign waveforms, we extend S5~\cite{S5}, a variant of SSMs~\cite{S4, Mamba} (Fig.~\ref{fig:model_arch}(b)). 
SSMs, inspired by state space representations in control theory~\cite{KalmanFilter}, have recently emerged as powerful architectures for long-sequence modeling~\cite{S4}, making them a natural choice for capturing the temporal dependencies in physiological signals.
In S5, the input sequence $\tilde{\bm{x}}_{t,k}\in\mathbb{R}^{n}$ is mapped to the output sequence $\bm{y}_{t,k}\in\mathbb{R}^{n}$ through latent states $\bm{h}_{t,k}\in\mathbb{R}^{m}$ as follows:
\begin{align}
    \bm{h}_{t,k} = \bar{\bm{A}}\bm{h}_{t,k-1} + \bar{\bm{B}}\tilde{\bm{x}}_{t,k},\quad \bm{y}_{t,k} = \bar{\bm{C}}\bm{h}_{t,k} + \bar{\bm{D}}\tilde{\bm{x}}_{t,k}\ ,
\end{align}
where $\tilde{\bm{x}}_{t,k}$ denotes the $k$-th timestep of the embedded sequence of $\bm{x}_t$, and $n$ is the embedding dimension while $m$ is the latent state dimension.
Moreover, $\bar{\bm{A}}\in\mathbb{R}^{m\times m}$ denotes the discretized state transition matrix, whereas $\bar{\bm{B}}\in\mathbb{R}^{m\times n}$, $\bar{\bm{C}}\in\mathbb{R}^{n\times m}$, and $\bar{\bm{D}}\in\mathbb{R}^{n\times n}$ denote the projection matrices.

\subsection{Model Architecture}
Fig.~\ref{fig:model_arch}(a) presents the architecture of the proposed method.
We extend S5 and introduce a dual-stream Flow-SSM block that jointly models PPG and vital signs, enabling fine-grained conditioning on PPG for vital sign reconstruction.
The architecture takes as input $\hat{\bm{x}}_t$, $\bm{z}$, and the timestep $t$ of the probability path in OT-CFM. 
After embedding $\hat{\bm{x}}_t$ and $\bm{z}$ with one-dimensional convolutional layers, the inputs are passed through a stack of $L$ Flow-SSM blocks to output $d\hat{\bm{x}}_t$.

Each Flow-SSM block adopts a dual-stream structure composed of LayerNorm~\cite{layernorm}, FiLM~\cite{FiLM}, S5, a scaling operation, and a feed-forward network.
FiLM and the scaling operation are employed for conditioning on $t$.
In FiLM, the scale parameter $\gamma$ and shift parameter $\beta$ of LayerNorm, and in the scaling operation, the scale parameter $\alpha$, are learnable parameters derived from the sinusoidal encoding of $t$, enabling feature distribution transformations that depend on $t$.
Furthermore, conditioning of $\hat{\bm{x}}_t$ on $\bm{z}$ is realized through an additive operation after a linear projection.
While cross-attention or post-concatenation mappings are commonly used for cross-modal conditioning~\cite{DiT}, in our setting $\hat{\bm{x}}_t$ and $\bm{z}$ are temporally aligned, making timestep-wise conditioning crucial (e.g., the systolic peak in PPG should closely align with the R-peak in ECG).
Therefore, we apply a simple additive operation at each timestep, enabling fine-grained, per-timestep conditioning of vital-sign reconstruction on PPG.
\section{Experiments}
\subsection{Datasets and Data Pre-processing}
We evaluate PENGUIN on six real-world PPG datasets across three vital sign reconstruction tasks (ECG reconstruction, respiratory monitoring, and ABP monitoring). 
Although the proposed framework is not limited to these tasks, following prior reviews~\cite{Naeini, PPGreview}, we focus on ECG, respiratory rate, and ABP, as these represent the primary targets in PPG-based vital sign reconstruction.
For each task, we selected datasets as follows: PPG-DaLiA~\cite{PPG-DaLiA} and WildPPG~\cite{WildPPG} for ECG reconstruction, BIDMC~\cite{BIDMC} and WESAD~\cite{WESAD} for respiratory monitoring, and UCI-BP~\cite{UCI-BP} and MIMIC-BP~\cite{MIMIC-BP} for ABP monitoring.

For pre-processing, we followed protocols in prior works \cite{RespDiff, RDDM, DeepCNAP}.
All signals were resampled to a uniform frequency of 128 Hz.
PPG was band-pass filtered with a Butterworth filter (0.5-4 Hz), then standardized using z-scores and scaled to the range [-1, 1].
ECG was high-pass filtered at 0.5 Hz using a Butterworth filter, followed by the same standardization and scaling procedure.
The respiratory rate was low-pass filtered at 1 Hz and likewise standardized and scaled.
Finally, for ABP, no further pre-processing was applied, as its amplitude carries critical physiological meaning (See \ref{metric}).

Across all datasets, we adopt a cross-subject paradigm, partitioning each dataset into training, validation, and test splits at a 6:1:1 ratio with no subject overlap.

\subsection{Evaluation Metrics}
\label{metric}
To assess the utility of the reconstructed vital signs, we employed well-established task-specific evaluation metrics following prior studies~\cite{RDDM, DeepCNAP, RespDiff}.
For ECG reconstruction, we computed the mean absolute error (MAE) of heart rate in beats per minute (bpm) between the reconstructed and ground-truth ECG, where heart rate was estimated from the ECG using the Hamilton method~\cite{Hamilton} with an 8-second window (hereafter denoted as HR Error).
For respiratory monitoring, we obtained the respiratory rate by applying a Fourier transform to both reconstructed and ground-truth respiratory waveforms, identifying the dominant non-negative frequency, and calculating the MAE in bpm with a 60-second window (hereafter RR Error).
For ABP monitoring, we evaluated the MAE (mmHg) of systolic and diastolic blood pressures over an 8-second window, corresponding to the maximum and minimum values of the ABP waveform, respectively (hereafter SBP Error and DBP Error).

\subsection{Implementation Details}
We trained the model with AdamW optimizer~\cite{adamw} ($\beta_1=0.9,\ \beta_2=0.999$) at a learning rate of $1.0\times 10^{-3}$, batch size 64, for up to 300 epochs, using early stopping with patience set to 10 epochs.
The hyperparameters were set to block number $L=4$, embedding dimension $n=128$, and latent state dimension $m=256$.
At inference, we use 25 sampling steps.

\begin{table*}[t]
\setlength{\tabcolsep}{5pt}
\centering
    \caption{Quantitative comparison across six PPG datasets. \textbf{Bold} and \underline{underlined} values denote the best and second-best performance, respectively. Parentheses indicate the performance margin between the proposed method and the best baseline.}
    \label{tab:main}
    \renewcommand{\arraystretch}{0.9}
    \begin{tabular}{l l c c c c c c}
    \toprule
    \multirow{2}{*}{Dataset}& 
    \multirow{2}{*}{Metric}& 
    \multicolumn{3}{c}{Specialist Model}& 
    \multicolumn{2}{c}{Generalist Model}\\ 
    \cmidrule(l{1mm}r{1mm}){3-5} 
    \cmidrule(l{1mm}r{1mm}){6-7} 
    && CycleGAN~\cite{CycleGAN}& RDDM~\cite{RDDM}& RespDiff~\cite{RespDiff}& 
    PaPaGei-S~\cite{PaPaGei}& \textbf{PENGUIN}\\
    \hline
    \rowcolor{gray!10}
    \multicolumn{7}{l}{\textit{\color{black!70}ECG Reconstruction}}\\
    PPG-DaLiA~\cite{PPG-DaLiA}& \multirow{2}{*}{HR Error~[bpm]}& 23.61& \underline{16.43}& 22.75& 40.89& \textbf{15.64}~({\color{ForestGreen}-0.79})\\
    WildPPG~\cite{WildPPG}& &23.21 &\underline{16.02}& 20.57& 55.42& \textbf{12.97}~({\color{ForestGreen} -3.05})\\
    
    \hline
    \rowcolor{gray!10}
    \multicolumn{7}{l}{\textit{\color{black!70}Respiratory Monitoring}}\\
    BIDMC~\cite{BIDMC}& \multirow{2}{*}{RR Error~[bpm]}& 9.78& 13.88& \underline{3.71}& 4.48& \textbf{2.98}~({\color{ForestGreen}-0.73})\\
    WESAD~\cite{WESAD}& & 11.93& 10.12& \underline{5.12}& 5.84& \textbf{4.45}~({\color{ForestGreen}-0.67})\\
    
    \hline
    \rowcolor{gray!10}
    \multicolumn{7}{l}{\textit{\color{black!70}ABP monitoring}}\\
    \multirow{2}{*}{UCI-BP~\cite{UCI-BP}}& SBP Error~[mmHg]& \underline{25.79}& 44.37& 78.83& 37.01& \textbf{12.61}~({\color{ForestGreen} -13.18})\\
    & DBP Error~[mmHg]& \underline{12.76}& 16.57& 26.13& 13.34& \textbf{7.14}~({\color{ForestGreen} -5.62})\\
    \multirow{2}{*}{MIMIC-BP~\cite{MIMIC-BP}}& SBP Error~[mmHg]& \underline{20.26}& 22.84& 97.65& 38.42& \textbf{17.43}~({\color{ForestGreen} -2.83})\\
    & DBP Error~[mmHg]& \textbf{10.49}& 11.83& 19.75& 11.52& \underline{11.34}~({\color{BrickRed} +0.85})\\
    \bottomrule
    \end{tabular}
\end{table*}
\begin{figure*}[t]
  \centering
  \includegraphics[width=\linewidth]{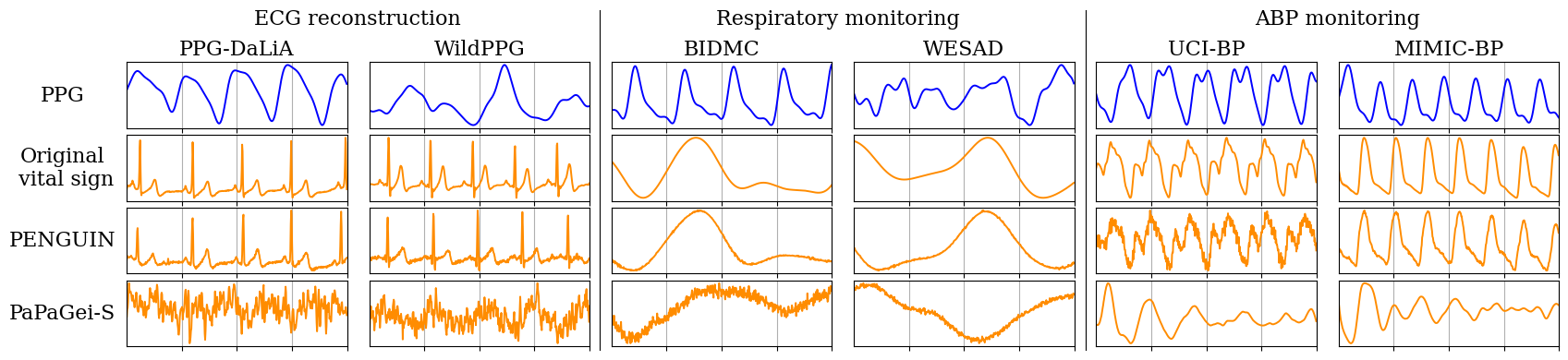}
  \caption{Qualitative comparison of reconstructed vital signs from PPG over a 4-second segment.}
  \label{fig:qual_result}
\end{figure*}

\subsection{Results and Discussion}
Table~\ref{tab:main} presents a quantitative comparison of the proposed method across six real-world PPG datasets.
As a direct baseline, we included PaPaGei-S~\cite{PaPaGei}, a recent PPG decoding method capable of handling multiple vital signs.
We further compared our method with CycleGAN~\cite{CycleGAN}, RDDM~\cite{RDDM}, and RespDiff~\cite{RespDiff}, which represent recent approaches tailored to specific vital sign reconstruction tasks from PPG. 
As shown in the table, the proposed method consistently outperformed all baselines in terms of HR error, RR error, SBP error, and DBP error, with the sole exception of DBP error on the MIMIC-BP dataset.
These findings suggest that the proposed method is broadly applicable to diverse PPG decoding scenarios, highlighting its potential for multiple vital sign reconstructions.

Fig.~\ref{fig:qual_result} illustrates a qualitative comparison of reconstructed vital signs from PPG over a 4-second segment.
The proposed method reliably preserves the morphological characteristics of the original vital signs, capturing sharp spikes in ECG reconstruction (i.e., QRS complex), reproducing respiratory rhythms that differ from PPG periodicity, and maintaining precise amplitude in ABP monitoring.

To further assess the proposed method, we conducted ablation studies as shown in Table~\ref{tab:ablation}.
We compared four configurations: (i) the proposed method; (ii) without FiLM conditioning; (iii) without Shift conditioning; and (iv) without PPG conditioning (i.e., synthesizing vital signs solely from noise).
On the PPG-DaLiA and WildPPG datasets, all Model variants (ii–iv) exhibited higher HR Error than Model (i), indicating that each conditioning contributes to performance improvement, with PPG conditioning exerting the greatest impact.
\section{Conclusion}
\setlength{\tabcolsep}{2pt}
\begin{table}[t]
    \centering
    \caption{Ablation study on conditioning strategies. \textbf{Bold} indicates the best performance.}
    \label{tab:ablation}
    \renewcommand{\arraystretch}{0.91}
    \begin{tabular}{c l c c c}
        \toprule
        \multicolumn{2}{l}{\multirow{2}{*}{\textbf{Model}}} & 
        PPG-DaLiA~\cite{PPG-DaLiA} & 
        WildPPG~\cite{WildPPG}\\
        \cmidrule(l{1mm}r{1mm}){3-3} 
        \cmidrule(l{1mm}r{1mm}){4-4} 
        & & \multicolumn{2}{c}{\makecell{HR Error~[bpm]}
        }\\
        \hline
        (i) & \textbf{PENGUIN}& \textbf{15.64}& \textbf{12.97}\\
        (ii) & w/o FiLM cond. &16.30 &13.24\\
        (iii) & w/o Shift cond. &15.72 &13.05 \\
        (iv) & w/o PPG cond. & 24.40& 21.75\\
        \bottomrule         
    \end{tabular}
\end{table}
In this study, we introduced PENGUIN, a generative flow-matching framework for reconstructing multiple vital signs as continuous waveforms from PPG. 
By extending deep state space models with a dual-stream Flow-SSM block, our method achieves fine-grained, per-timestep conditioning on PPG, thereby preserving the morphological fidelity of vital signs. 
Comprehensive evaluation across six real-world datasets covering ECG reconstruction, respiratory monitoring, and ABP monitoring demonstrated consistent improvements over both task-specific and general-purpose baselines. 
Future directions include validating the framework across a wider range of vital signs and enabling flexible windowing.

\ninept
\vfill\pagebreak
\bibliographystyle{IEEEbib}
\bibliography{refs}

\end{document}